# Estimation of the incubation period and generation time of SARS-CoV-2 Alpha and Delta variants from contact tracing data

Mattia Manica [a,*], Maria Litvinova [b,*], Alfredo De Bellis [a,c], Giorgio Guzzetta [a], Pamela Mancuso [d], Massimo Vicentini [d], Francesco Venturelli [e], Eufemia Bisaccia [e], Ana I. Bento [b], Piero Poletti [a], Valentina Marziano [a], Agnese Zardini [a], Valeria d'Andrea [a], Filippo Trentini [a,f], Antonino Bella [g], Flavia Riccardo [g], Patrizio Pezzotti [g], Marco Ajelli [b,#], Paolo Giorgi Rossi [d,#], Stefano Merler [a,#,%] and the Reggio Emilia COVID-19 Working Group.

  a. Center for Health Emergencies, Fondazione Bruno Kessler, Trento, Italy
  b. Laboratory for Computational Epidemiology and Public Health, Department of Epidemiology and Biostatistics, Indiana University School of Public Health, Bloomington, IN, USA
  c. Department of Mathematics, University of Trento, Trento, Italy
  d. Epidemiology Unit, Azienda Unità Sanitaria Locale – IRCCS di Reggio Emilia, Reggio Emilia, Italy
  e. Public Health Department, Azienda Unità Sanitaria Locale – IRCCS di Reggio Emilia, Reggio Emilia, Italy
  f. Dondena Centre for Research on Social Dynamics and Public Policy, Bocconi University, Milan, Italy
  g. Dipartimento di Malattie Infettive, Istituto Superiore di Sanità, Rome, Italy

[*] joint first authors
[#] joint senior authors
[%] corresponding author: merler@fbk.eu

The following are members of the Reggio Emilia Covid-19 Working Group: Emanuela Bedeschi, Cinzia Perilli, Nadia Montanari, Francesca Pia Leonetti, Nicoletta Patrignani, Letizia Bartolini, Francesca Roncaglia, Isabella Bisceglia, Valeria Cenacchi, Barbara Braghiroli, Anna Pezzarossi

**Summary**

Background. During 2021, the COVID-19 pandemic was characterized by the emergence of lineages with increased fitness. For most of these variants, quantitative information is scarce on epidemiological quantities such as the incubation period and generation time, which are critical for both public health decisions and scientific research.

Method. We analyzed a dataset collected during contact tracing activities in the province of Reggio Emilia, Italy, throughout 2021. We determined the distributions of the incubation period using information on negative PCR tests and the date of last exposure from 282 symptomatic cases. We estimated the distributions of the intrinsic generation time (the time between the infection dates of an infector and its secondary cases under a fully susceptible population) using a Bayesian inference approach applied to 4,435 SARS-CoV-2 cases clustered in 1,430 households where at least one secondary case was recorded.

Results. We estimated a mean incubation period of 4.9 days (95% credible intervals, CrI, 4.4-5.4; 95 percentile of the mean distribution: 1-12) for Alpha and 4.5 days (95%CrI 4.0-5.0; 95 percentile: 1-10) for Delta. The intrinsic generation time was estimated to have a mean of 6.0 days (95% CrI 5.6-6.4; 95 percentile: 1-15) for Alpha and of 6.6 days (95%CrI 6.0-7.3; 95 percentile: 1-18) for Delta. The household serial interval was 2.6 days (95%CrI 2.4-2.7) for Alpha and 2.4 days (95%CrI 2.2-2.6) for Delta, and the estimated proportion of pre-symptomatic transmission was 54-55% for both variants.

Conclusions. These results indicate limited differences in the incubation period and intrinsic generation time of SARS-CoV-2 variants Alpha and Delta compared to ancestral lineages.

Funding. EU grant 874850 MOOD

**Introduction**

The second year of the COVID-19 pandemic has been characterized by the global emergence of several lineages which were able to replace circulating ones thanks to their increased fitness [1]. In particular, 2021 saw the sequential rise and fall of two variants of concern, Alpha and Delta, the latter of which has been rapidly outpaced by Omicron around the end of 2021. Compared to ancestral strains, scarce quantitative information is available on several variant-specific epidemiological quantities, among which the incubation period (i.e., the time elapsed between infection episode and symptom onset) and the generation time (i.e., the time elapsed between the infection episode of a primary case and that of a secondary case). These two quantities are especially important to define the duration of isolation for infectious individuals and of quarantines for close contacts and travelers, as well as protocols for community-based interventions such as contact tracing activities [2-4] and class/school closures [3, 5]. The knowledge of the generation time distribution also informs the estimation of the net reproduction number (i.e., the average number of new cases generated by an infectious case at a given time of the epidemic) [6], which is a key indicator for monitoring epidemic outbreaks and defining population-level measures, such as physical distancing and movement restrictions [7].

The incubation time is mostly a biologically determined parameter since it depends on virus characteristics and virus-host immunological and pathological interactions. On the other hand, the generation times that occur in a population depend on the interactions between infectious individuals and their contacts, and therefore may be subject to specific epidemiological conditions in which they are measured, including individual behaviors, environmental determinants, and control measures put in place [8]. For example, the generation time realized in households is remarkably shortened with respect to the one observed in the general community due to the depletion of susceptible individuals and the competition of simultaneously infectious individuals to find susceptible household members to infect [8]. In contrast to the "realized" distribution of the generation time, occurring in a realistic network of contacts, the "intrinsic" generation time represents the generation time that would be observed in a fully susceptible, homogenously mixed population [9]. The intrinsic generation time is therefore less dependent on the specific conditions of the epidemiological setting from which it is inferred. In this study, we estimate the distribution of incubation periods and generation times (both intrinsic and realized) for SARS-CoV-2 variants Alpha and Delta, using a Bayesian inference approach applied to data collated during COVID-19 contact tracing activities in the province of Reggio Emilia, Italy, during 2021.

**Table 1. Descriptive statistics of SARS-CoV-2 cases in the household datasets for Alpha and Delta variants.**

|  | **ALPHA** | **DELTA** |
|---|---|---|
| **Period** | March 1 – April 30, 2021 | August 1 – October 31, 2021 |
| **Number of cases** | 2,988 | 1,447 |
| **Clinical outcome (%):** | | |
|    Symptomatic | 1,822 (61.0%) | 1,137 (78.6%) |
|    Asymptomatic | 1,166 (39.0%) | 310 (21.4%) |
| **Gender (%):** | | |
|    Male (%) | 1,447 (48.4%) | 702 (48.5%) |
|    Female (%) | 1,541 (51.6%) | 745 (51.5%) |
| **Age group (%):** | | |
|    0-15 years old | 978 (32.7%) | 432 (29.8%) |
|    16-44 years old | 1,090 (36.5%) | 529 (36.6%) |
|    45-64 years old | 734 (24.6%) | 329 (22.7%) |
|    65+ years old | 186 (6.2%) | 157 (10.9%) |
| **Vaccination status at the end of the period (%):** | | |
|    1 dose | 71 (2.4%) | 142 (9.8%) |
|    2 doses | 22 (0.7%) | 444 (30.7%) |
|    3 doses | 0 (0.0%) | 2 (0.1%) |
|    None | 2895 (96.9%) | 859 (59.4%) |
| **Number of households** | 906 | 524 |
| **Average household size (95% quantile)** | 3.5 (2 – 6) | 2.96 (2 – 7) |

**Results**

For the estimation of the generation time, we considered SARS-CoV-2 transmission in households, and statistics of the corresponding dataset are described in Table 1. Figure 1 shows a schematization of an illustrative household cluster, with the corresponding dates of infection, symptom onset, diagnosis and negative tests for individuals, as well as relevant intervals to be estimated.

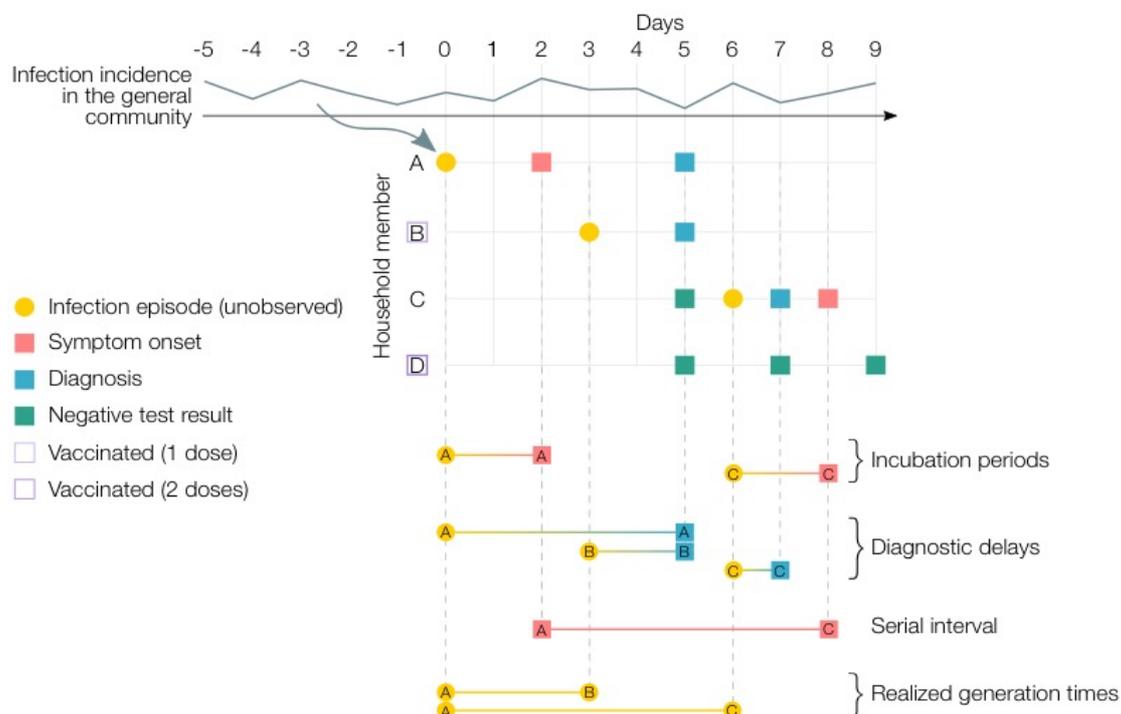

**Figure 1. Illustrative example of a household cluster.** A household with 4 members, of which A was infected outside the household (in the general community) at day 0 and then transmitted to cases B (asymptomatic) and C (symptomatic), while D remained uninfected. B and D were vaccinated with 1 and 2 doses respectively. A hypothetical epidemic curve in the general community, representing the external force of infection on household members, is reported on top of the graph. Circles indicate unobserved events; squares indicate observed events. Examples of the temporal intervals of interest for the estimates of this work are reported in the bottom part of the figure. Note that for the household serial interval and the realized household generation time, the source of infection (whether from outside the household or from a household member, and, in the latter case, which household member) is also unobserved and needs to be probabilistically reconstructed. The intrinsic generation time is not displayed as it represents the distribution of generation times among infections occurring in the general population in a fully susceptible population [9].

The best fit for the distributions of the incubation period was a gamma distribution, with a mean of 4.9 days for Alpha (95% Credible Intervals of the mean, CrI, 4.4-5.4; 95% quantile of the mean distribution 1-12 days) and of 4.5 days for Delta (95% CrI, 4.0-5.0; 95% quantile of the mean distribution 1-10 days) (Figure 2 and Table 2). The differences between empirical distributions of incubation periods for Alpha and Delta variants were not statistically significant (Wilcoxon-type test p-value 0.45). Unsurprisingly, the estimate for the incubation period was longer (mean: 7.3-7.4 days for Alpha and 6.2-6.3 days for Delta) and had a larger uncertainty when including in the estimation those cases for which the date of earliest exposure was unknown (Appendix).

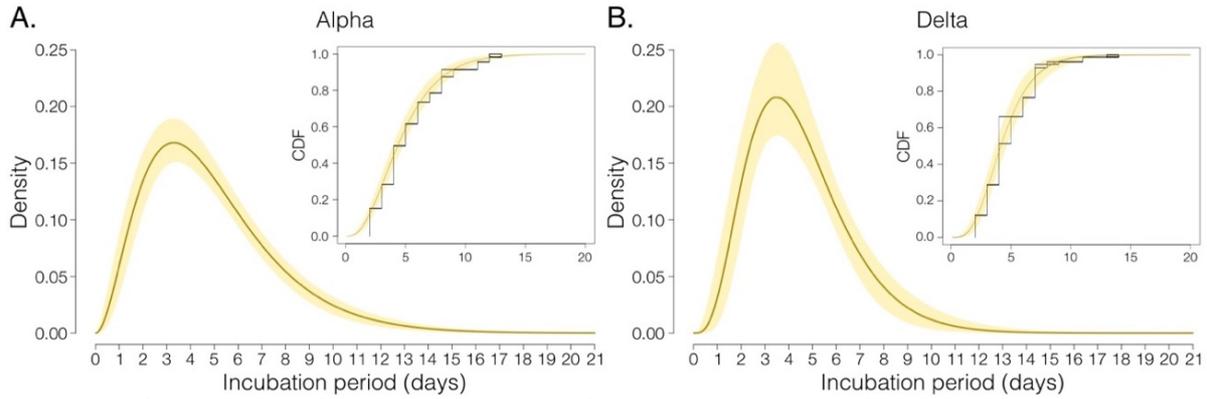

**Figure 2. Estimation of the incubation period for the Alpha and Delta SARS-CoV-2 variants.** A) Probability density function (PDF) of the estimated distribution of incubation period for Alpha variant with 95% credible intervals (CrI) based on nonparametric bootstrap resampling of the distribution parameters (10,000 samples). Line: mean PDF; shaded area: bootstrapped pointwise 95% credible intervals (CrI). The inset shows the cumulative distribution function (CDF) of the empirical distribution (black line) where rectangles represent areas of non-unique empirical distribution function, CDF of the distribution fitted to interval-censored data (line) and bootstrapped pointwise 95% CrI on probabilities (shaded area). B) as A, but for Delta variant.

**Table 2. Estimates for the incubation period, diagnostic delay, intrinsic and realized generation time, and household serial intervals.** Reported parameters of shape and scale for the incubation period and intrinsic generation time refer to a gamma distribution.

| | | ALPHA | DELTA |
|---|---|---|---|
| **INCUBATION PERIOD** | mean (95%CrI) [days] | 4.9 (4.4-5.4) | 4.5 (4.0-5.0) |
| | 95% quantile of the mean distribution [days] | 1-12 | 1-10 |
| | shape mean (95%CrI) | 3.08 (2.56-3.86) | 4.43 (3.26-6.70) |
| | scale mean (95%CrI) | 1.58 (1.24-1.93) | 1.01 (0.65-1.43) |
| **DIAGNOSTIC DELAY** | mean (95% quantile) [days] | 7.2 (2-14) | 6.9 (2-14) |
| **INTRINSIC GENERATION TIME** | mean (95%CrI) [days] | 5.95 (5.57-6.44) | 6.62 (6.01-7.31) |
| | 95% quantile of the mean distribution [days] | 1-15 | 1-18 |
| | shape mean (95%CrI) | 2.56 (2.35-2.79) | 2.28 (2.04-2.59) |
| | scale mean (95%CrI) | 2.33 (2.06-2.61) | 2.91 (2.24-3.51) |
| **REALIZED HOUSEHOLD GENERATION TIME** | mean (95%CrI) [days] | 4.08 (4.06-4.09) | 3.96 (3.94-4.01) |
| **HOUSEHOLD SERIAL INTERVAL** | mean (95%CrI) [days] | 2.40 (2.20-2.60) | 2.56 (2.37-2.74) |
| **PRE-SYMPTOMATIC TRANSMISSION** | Mean (95%CrI) [%] | 54.4 (51.9-56.8) | 54.5 (51.8-56.8) |

The resulting estimated distribution of delays between infection and diagnosis (used to assign infection dates for asymptomatic individuals) had a mean of 7.2 days (95% quantile: 2-14 days) for the Alpha variant (Table 2). The mean intrinsic generation time estimated for Alpha was 5.95 days (95% CrI of the mean: 5.57-6.44 days) and the mean realized household generation time was 4.08 days (95%CrI of the mean: 4.06-4.09 days) (Figure 3 and Table 2). The mean household serial interval was 2.40 days (95%CrI of the mean: 2.20-2.60 days), with 54.4% (95%CrI: 51.9-56.8%) of transmission episodes being pre-symptomatic (i.e., secondary cases transmitted by cases who would develop symptoms after the transmission episode). Sensitivity analyses yielded similar results, with the mean intrinsic generation time ranging between 5.42 and 6.71 days, the mean realized household generation time ranging between 3.84 and 4.71 days, and the mean household serial interval ranging between 2.16 and 2.54 days (Appendix).

The estimated distribution of delays between infection and diagnosis had a mean of 6.9 days for the Delta variant (95% quantile: 3-15 days) (Table 2). The mean intrinsic generation time estimated for Delta was 6.62 days (95%CrI of the mean: 6.01-7.31 days) and the mean realized household generation time was 3.96 days (95%CrI of the mean 3.94-4.01 days) (Figure 3 and Table 2). The mean household serial interval was 2.56 days (95%CrI of the mean 2.37-2.74 days), with 54.5% (95%CrI: 51.8-56.8%) of transmission episodes being pre-symptomatic. Sensitivity analyses yielded similar results, with the mean intrinsic generation time ranging between 6.40 and 7.76 days, the mean realized household generation time ranging between 3.89 and 4.55 days, and the mean household serial interval ranging between 2.13 and 2.59 days (Appendix).

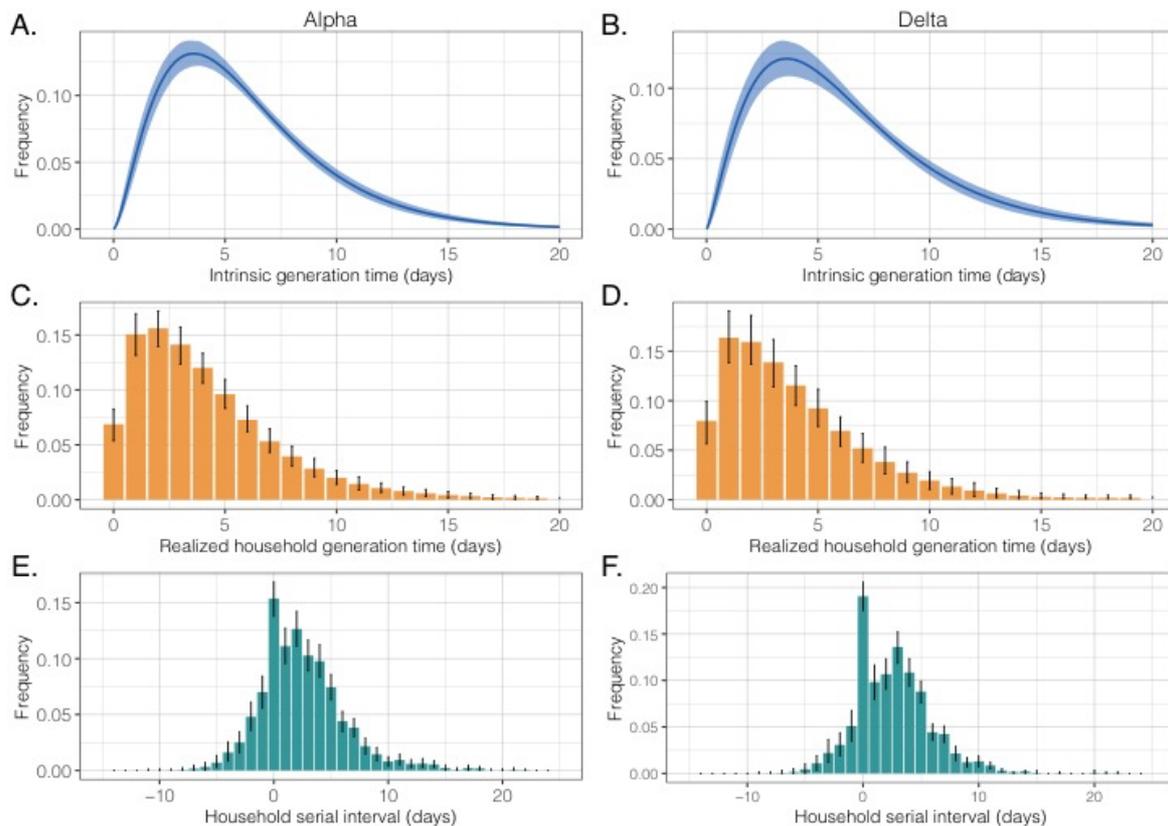

**Figure 3. Estimates of generation times and household serial intervals for the Alpha and Delta variants.** A) Distribution of the intrinsic generation time for the Alpha variant; solid line: mean estimate; shaded area: 95% CrI; B) As A, but for Delta; C) Distribution of the realized household generation time for the Alpha variant; bars: mean estimate; vertical lines: 95% CrI; D) As C, but for Delta; E) Distribution of the household serial interval for the Alpha variant; bars: mean estimate; vertical lines: 95%CrI; F) As E, but for Delta.

**Discussion**

We estimated the distribution of the incubation period and generation time for SARS-CoV-2 Alpha and Delta variants by analyzing comprehensive data collected during contact tracing activities in the province of Reggio Emilia, Italy, throughout 2021. We found no statistical difference for the duration of the incubation period for Alpha (mean: 4.9 days) and Delta (mean: 4.5 days) variants. Both estimates are in line (albeit slightly shorter) with those obtained for the ancestral lineage [11-14] and in the range of those obtained for variants [15-21]. Our estimates of the mean generation time for both Alpha (mean: 5.95 days) and Delta (mean: 6.62 days) are compatible with previous estimates for ancestral lineages [22-26], including a previous estimate for Italy of 6.68 days [27]. We also found comparable household serial intervals between Alpha (mean: 2.40 days) and Delta (mean: 2.56 days) with similar proportions of pre-symptomatic transmissions (about 55% for both variants). Previous estimates on both ancestral lineages and Alpha and Delta variants are highly variable (due to the high sensitivity of these parameters to epidemiological conditions of the study settings) and ranged between 1.8 days and 7.5 days [11-12, 28-30] for the serial interval and between 13% and 65% [11, 13, 28] for the proportion of pre-symptomatic transmission.

Estimates of the intrinsic generation time may still depend on epidemiological specificities of the geographical setting from which the data are collected, as well as by the inference method. This may explain why our estimate of the mean intrinsic generation time for the Alpha variant substantially overlaps with estimates from England [29], while our estimates for the Delta variant are substantially longer (mean: 6.62 days vs. 4.7 days). In our dataset, significantly different epidemiological conditions across the two study periods were observed. The study location was under tight restrictions, including stay-at-home mandates, for a large portion of the period selected for Alpha (March 1 – April 30); furthermore, the vaccination program was still taking off (two-dose coverage between 2 and 10% of the total Italian population) and cold temperatures likely resulted in poor room ventilation inside households. On the other hand, during the period selected for Delta (August 1 – October 31) there was low viral circulation and most restrictions were lifted, possibly thanks to the expansion of the vaccination coverage (rising from 53 to 72% of the total population over the period) and to higher ventilation and less time spent indoor, afforded by warmer temperatures in the summer and early fall. More in general, given its potential sensitivity to local factors, we point out the need to obtain country-specific estimates of the distribution of the generation time. For what concerns Italy, this study suggests the adequacy of epidemiological analyses (i.e., computation of reproduction numbers; modelling estimates) performed by assuming a distribution of the generation time similar to ancestral lineages.

A main strength of this work consists in the very large population-based dataset that comprehensively covers household clusters observed in the province of Reggio Emilia. The tracing and testing protocol was identical in the two periods and public health officials made efforts to have high compliance to testing policies (97% of individuals who were offered a test accepted at least once), including testing all household members of cases at the date of the first diagnosis in the household. To minimize the possibility that our data contain clusters due to other variants, we selected two periods where Alpha and Delta were largely dominant [11]. However, for the Alpha period a residual circulation (7-8% prevalence) of the Gamma variant was detected in the Emilia-Romagna region [10, 31]. The estimates of the intrinsic generation time can be compared across periods with different vaccination coverage since the model includes susceptibility and transmissibility variations according to the individual's vaccination history. A limitation of the model is its reliance on assumptions for the dates of infection of infected individuals; nonetheless, estimates were substantially robust with respect to different methods of imputation and different distributions of the incubation period (Appendix). The same intrinsic limitation of the unobservability of infection times is shared by all transmission chain reconstruction models, but there are now several examples where these models have been proven to correctly identify the transmission dynamics of infectious outbreaks [32-36]. A specific limitation of this study was the lack of information about previous SARS-CoV-2 infection in undiagnosed individuals. In the main analysis we assumed that all undiagnosed individuals did not have a pre-existing protection from natural immunity. However, in a sensitivity analysis, we show that assuming full protection from previous infection in a fraction of undiagnosed individuals hardly affects our results (Appendix). Another specific limitation is that we assumed 100% compliance to quarantine protocols (i.e., that household members quarantined after diagnosis of another member could only be infected within the household). A sensitivity analysis where quarantines of household members are not considered (i.e., 0% compliance) yielded similar results to the ones illustrated in the main analysis (Appendix).

**Conclusion**

Results from this study suggest that the length of the incubation period and generation time for Alpha and Delta variants are comparable to that of the ancestral lineage. These findings provide support to the recommendations of adopting duration of quarantine, isolation, and contact tracing operations similar to those for the ancestral lineage. This work also establishes a method for rapidly estimating the incubation periods and generation times on further emerging variants of concern, provided that high-quality contact tracing data are made available.

**Methods**

*Data*
To mitigate the spread of SARS-CoV-2, contact tracing activities were carried out in the province of Reggio Emilia, Italy, throughout the duration of the pandemic. Identified SARS-CoV-2 cases occurring in the province were confirmed via a Polymerase Chain Reaction (PCR) assay, reported in real time to the public health service of the Reggio Emilia local health authority and isolated at home until a negative PCR test result and for a maximum of 21 days. During the study period all antigenic positive tests were confirmed with PCR. All cases were contacted via telephone to identify their close contacts. A close contact was defined as a person who stayed in the same room with a confirmed case without a face mask, or for more than 15 minutes at less than 2 meters, between 2 days before to 10 days after symptom onset (for symptomatic cases) or diagnosis (for asymptomatic infections). Contacts were tested and quarantined at home for 10 days, if they had a negative PCR test result at that date, or for 14 days without testing [37]. All household members of a case were quarantined until a negative test after the end of the isolation period for the index case. Compliance with at least one of the tests proposed by the public health service was 97.0% during the study period (March-October 2021).
Data on test results, symptom onset date (if applicable), and setting of likely transmission were collected for all identified cases and their contacts and were linked to individual records on vaccination history (first, second, and booster doses), as well as the date of the last reported contact with a known case (date of last exposure). Appropriate data quality checks were conducted in strict collaboration with the Reggio Emilia local health authority to minimize missing information and accurately define household clusters. A household cluster was defined as households with at least two positive individuals with a diagnosis spaced less than 25 days apart.
Since genomic information on the variant was not available, we conservatively defined two time periods where circulation of SARS-CoV-2 in the Region was almost exclusively attributable (at least ~90% prevalence) to variant Alpha (March 1 – April 30, 2021) and to Delta (August 1 – October 31, 2021) [10].

*Ethics*
Covid-19 surveillance data collection, including specific studies on contact tracing, was authorized by the Italian Presidency of the Council of Ministers on 27 February 2020 (ordinance No 640, article 1 [38]).

*Estimation of the incubation period*
For the estimation of the incubation period, we selected all symptomatic cases with a date of diagnosis within either of the two periods defined for Alpha or Delta. For each case, the date of the last negative PCR test $T_N$ and the date of last exposure $T_L$ were used to set the limits for the earliest and last exposure, respectively. Therefore, we excluded all cases for which either date was unavailable or for which the condition $T_N \leq T_L \leq T_S$, where $T_S$ is the date of symptom onset, did not hold. The resulting sample for the estimation contains 193 observations for Alpha and 89 for Delta. We used the generalization of the Wilcoxon-Mann-Whitney test for interval-censored data to compare the empirical data in the two samples. Two parametric distributions (gamma and Weibull) were fit to the interval-censored empirical data on the time between likely infection and the symptom onset [11, 28] using a maximum likelihood optimization. The best fit was selected based on the minimum Akaike information criterion [39]. Credible Intervals (CrI) for the parameters of estimated distributions were obtained from the 95% percentile of estimates over 10,000 bootstrap samples for censored data. See Appendix for further details on the method and for sensitivity analyses on estimation criteria.

*Estimation of the generation time and of the serial intervals*
For the estimation of the generation time, we selected only household clusters for which all dates of diagnosis were included in either of the two periods defined for Alpha or for Delta. To reduce the possibility of missed diagnoses in the households due to false negative test results, we further selected households for which undiagnosed members had at least two negative test results. We extended a Bayesian inference model for the reconstruction of transmission links in households [40]. The model exploits the temporal information on SARS-CoV-2 infections recorded in the dataset to probabilistically identify, for every case, the likely source of infection (from outside the household or from a specific household member). Parameters for the generation time, which we assume to be Gamma-distributed, are simultaneously calibrated via a Markov Chain Monte Carlo approach where the likelihood of the observed data is defined mechanistically through the computation of the force of infection to which all individuals are subject over time. The force of infection includes information on the temporal incidence of cases in the general population, and on the date of infection and vaccination history for any individual. For each symptomatic case, the date of infection was imputed by subtracting the time of symptom onset by a randomly sampled incubation period from the estimated variant-specific distribution. The imputed dates of infection for symptomatic individuals defined a distribution of delays between infection and diagnosis (diagnostic delay distribution), which was used to impute the date of infection of asymptomatic individuals starting from their date of diagnosis. For both symptomatic and asymptomatic individuals, we weighted the probability of an imputed date of infection using the available information on previous negative test results, allowing for the possibility of false negatives. The sampling of infection dates was repeated 100 times and the Bayesian model was re-calibrated on each resampling. CrI for the estimated parameters were obtained from the 95% percentile of the resulting pooled distributions.

The algorithm described above assumes that symptomatic and asymptomatic individuals have the same distribution of diagnostic delays; because we could not test this hypothesis, we considered in a sensitivity analysis an alternative method for inferring the date of infection of asymptomatic individuals. As further sensitivity analyses, we imputed dates of infection using two alternative distributions of the incubation period previously estimated for ancestral SARS-CoV-2 lineages [12, 28], and an alternative model where we assume a halved transmissibility for asymptomatic individuals [41]. A full description of the Bayesian inference model is available in the Appendix.

The adopted Bayesian approach allowed us to estimate both the intrinsic generation time and the realized household generation time. We also estimated the distribution of the household serial interval from the differences of symptom onset dates in each infector-infectee pair inferred by the model where both are symptomatic.

## Author contributions

MM, ML, GG, MA, and SM conceived the study. MM, ML, AdB and GG wrote the first draft of the manuscript. MM, ML and AdB wrote the code and performed the analyses. PM, MV, FV, EB, AB, FR, and the members of the Reggio Emilia COVID-19 Working Group collected the epidemiological data. MM, ML, AdB, GG, AIB, VM, PPo, AZ, VdA, FT, MA, PGR, SM interpreted results. GG, PPe, MA, PGR and SM supervised the study. All authors read, reviewed, and approved the manuscript for submission.

## Data availability statement

Data and code for reproducing our results will be made available on a public repository upon review.

## Acknowledgments

GG, SM, MM and AZ acknowledge funding from EU grant 874850 MOOD (catalogued as MOOD 000)

## Conflict of interest

MA has received research funding from Seqirus. The funding is not related to COVID-19. All other authors declare no competing interest.

# Appendix

Estimation of the incubation period and generation time of SARS-CoV-2 Alpha and Delta variants from contact tracing data

Mattia Manica [a,*], Maria Litvinova [b,*], Alfredo De Bellis [a,c], Giorgio Guzzetta [a], Pamela Mancuso [d], Massimo Vicentini [d], Francesco Venturelli [e], Eufemia Bisaccia [e], Ana I. Bento [b], Piero Poletti [a], Valentina Marziano [a], Agnese Zardini [a], Valeria d'Andrea [a], Filippo Trentini [a,f], Antonino Bella [g], Flavia Riccardo [g], Patrizio Pezzotti [g], Marco Ajelli [b,#], Paolo Giorgi Rossi [d,#], Stefano Merler [a,#,%] and the Reggio Emilia COVID-19 Working Group.

- h. Center for Health Emergencies, Fondazione Bruno Kessler, Trento, Italy
- i. Laboratory for Computational Epidemiology and Public Health, Department of Epidemiology and Biostatistics, Indiana University School of Public Health, Bloomington, IN, USA
- j. Department of Mathematics, University of Trento, Trento, Italy
- k. Epidemiology Unit, Azienda Unità Sanitaria Locale – IRCCS di Reggio Emilia, Reggio Emilia, Italy
- l. Public Health Department, Azienda Unità Sanitaria Locale – IRCCS di Reggio Emilia, Reggio Emilia, Italy
- m. Dondena Centre for Research on Social Dynamics and Public Policy, Bocconi University, Milan, Italy
- n. Dipartimento di Malattie Infettive, Istituto Superiore di Sanità, Rome, Italy

[*] joint first authors
[#] joint senior authors
[%] corresponding author: merler@fbk.eu

The following are members of the Reggio Emilia Covid-19 Working Group: Emanuela Bedeschi, Cinzia Perilli, Nadia Montanari, Francesca Pia Leonetti, Nicoletta Patrignani, Letizia Bartolini, Francesca Roncaglia, Isabella Bisceglia, Valeria Cenacchi, Barbara Braghiroli, Anna Pezzarossi

# Table of Contents

# 1. Estimation of the incubation period

For the estimation of the incubation period, we considered observations on individuals who were contacts of an index case and who later became symptomatic and diagnosed with SARS-CoV-2. Among these, we selected cases having a diagnosis in either of the selected study period for Alpha (1021 cases) or Delta (519 cases). The date of symptom onset and the date of last exposure was available for all symptomatic cases. For each case, the potential incubation period was bounded by the date of the latest negative test result before the diagnosis (earliest possible exposure) and by the date of the last exposure. We excluded 172 Alpha cases and 53 Delta cases for which the information on exposure were conflicting (e.g., last negative test successive to the last reported exposure), 298 Alpha cases and 173 Delta cases for which the date of last exposure was successive to symptom onset, and 358 Alpha cases and 204 Delta cases for which only a date of last exposure was available, obtaining 193 Alpha cases and 89 Delta cases for the main analysis (see Figure S1 for the sample selection). The resulting censored intervals of the possible incubation periods are reported for all cases in Figure S2.

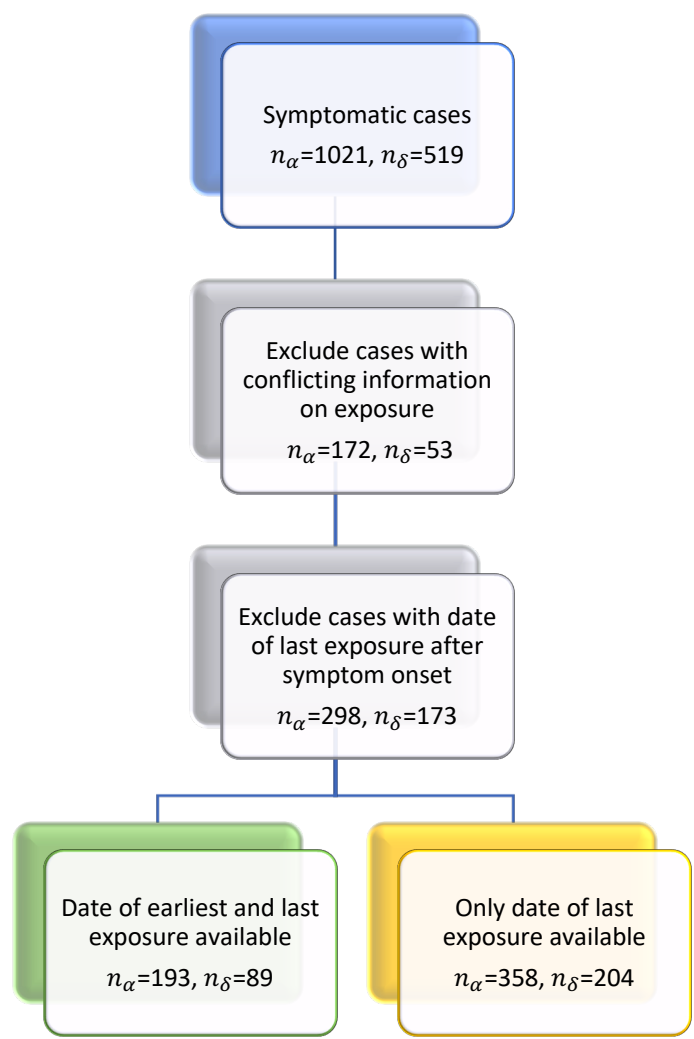

**Figure S1. Workflow of sample selection.** Gray boxes represent exclusion steps. The green box shows the sample used for the main analysis. The yellow box shows the additional sample used for a sensitivity analysis. $n_\alpha$ represents the sample size for Alpha variant, $n_\delta$ represents the sample size for Delta variant

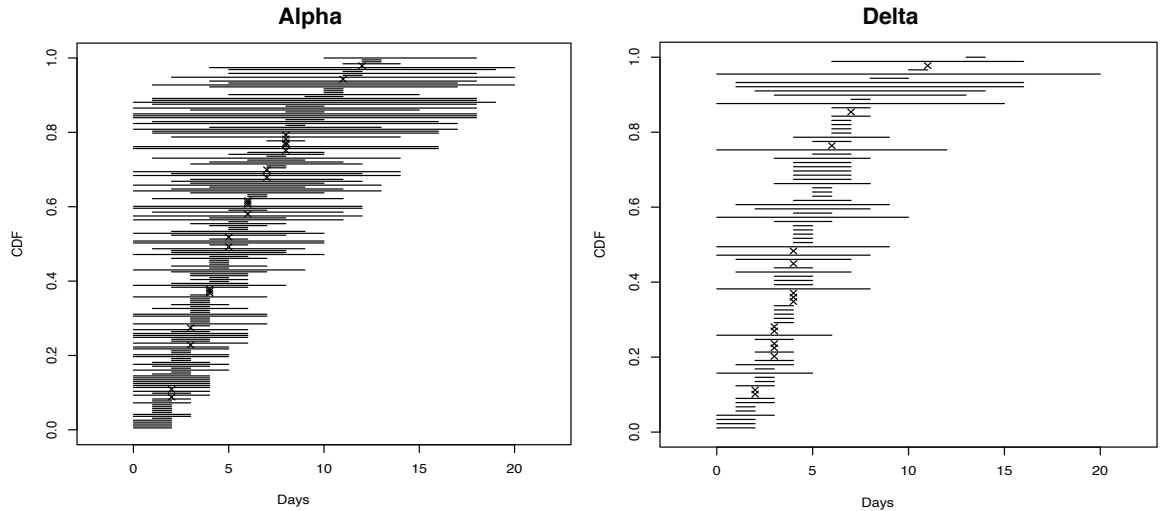

**Figure S2. Incubation period censored data.** Interval censored and non-censored observations for each case are ordered by their mid-points.

We estimated both Gamma and Weibull distributions using the censored data. Maximum likelihood estimations of the distribution parameters were calculated by using the *fitdistrplus* package in R. Direct optimization of the log-likelihood is performed using general-purpose optimization based on Nelder–Mead, quasi-Newton algorithm for both Gamma and Weibull distributions. Nonparametric bootstrap resampling was used to simulate uncertainty in the parameters of the estimated distributions. Results of the estimation procedure described in the main text are presented in Table S1.

**Table S1. Estimated distribution of the incubation period.** SD: Standard Deviation. AIC: Akaike Information Criterion

| Variant | Distribution | Parameters: mean (SD) | Mean distribution (days) | | | AIC score |
|---|---|---|---|---|---|---|
| | | | Mean | SD | 95% percentile | |
| *Alpha* (N=193) | Gamma | shape = 3.08 (0.39), rate = 0.63 (0.084) | 4.9 | 2.8 | 1.0 – 11.7 | 506.9 |
| | Weibull | scale = 5.52 (0.27), shape = 1.83 (0.13) | 4.9 | 2.8 | 0.7 – 11.3 | 510.7 |
| *Delta* (N=89) | Gamma | shape = 4.43 (0.76), rate = 0.99 (0.18) | 4.5 | 2.1 | 1.3 - 9.6 | 261.3 |
| | Weibull | scale = 5.09 (0.30), shape = 2.10 (0.18) | 4.5 | 2.2 | 0.9 - 9.5 | 267.1 |

*Sensitivity analyses*
As a first sensitivity analysis, we added the observations with only the date of last exposure being available (see Figure S1). For cases with unknown date of earliest possible exposure, we set the maximum boundary of the incubation period to 21 days before the symptom onset. Figure S3 shows the censored data used in this estimation. This increased both the sample size and the uncertainty regarding the earliest possible exposure, naturally increasing the average of the estimated incubation period (Table S2). Then, we repeated the main analysis and the sensitivity analysis above after selecting cases falling within the Alpha or Delta period based on the date of symptom onset rather than on the date of diagnosis, obtaining similar results to the corresponding analyses above.

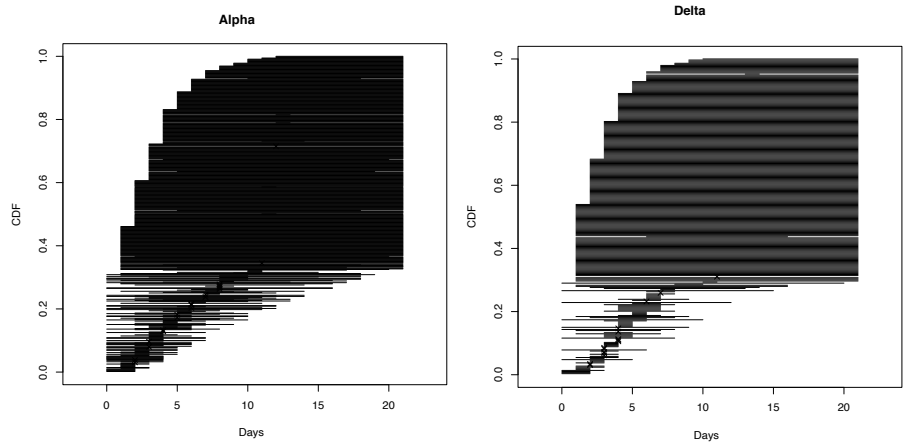

**Figure S3. Incubation period censored data, including observations with no information on earliest possible exposure (sensitivity analysis A).** Interval censored and non-censored observations for each case are ordered by their mid-points.

**Table S2. Estimated distribution of incubation period in sensitivity analyses.** SD: Standard Deviation. AIC: Akaike Information Criterion

| Variant | Distribution | Parameters: mean (SD) | Mean distribution (days) | | | AIC score |
|---|---|---|---|---|---|---|
| | | | Mean | SD | 95% percentile | |
| A) Date of the last exposure available (independently from availability of earliest exposure) | | | | | | |
| *Alpha (N=551)* | Gamma | shape = 3.75 (0.31), rate = 0.51 (0.05) | 7.3 | 3.8 | 1.9 - 16.6 | 770.9 |
| | Weibull | scale = 8.33 (0.29), shape = 2.23 (0.12) | 7.4 | 3.5 | 1.6 - 15.0 | 764.7 |
| *Delta (N=293)* | Gamma | shape = 4.58 (0.60), rate = 0.73 (0.12) | 6.3 | 2.9 | 2.0 - 13.2 | 376 |
| | Weibull | scale = 7.07 (0.33), shape = 2.49 (0.19) | 6.3 | 2.7 | 1.7 - 12.0 | 380.2 |
| B) As main analysis, but cases are assigned to variant via date of symptom onset | | | | | | |
| *Alpha (N=187)* | Gamma | shape = 3.12 (0.40), rate = 0.65 (0.087) | 4.8 | 2.7 | 1.0 - 11.4 | 495.5 |
| | Weibull | scale = 5.45 (0.27), shape = 1.84 (0.13) | 4.8 | 2.7 | 0.7 - 11.1 | 499.2 |
| *Delta (N=89)* | Gamma | shape = 4.70 (0.81), rate = 1.04 (0.19) | 4.5 | 2.1 | 1.4 - 9.4 | 255.9 |
| | Weibull | scale = 5.10 (0.30), shape = 2.14 (0.19) | 4.5 | 2.2 | 0.9 - 9.3 | 263 |
| C) As sensitivity analysis A), but cases are assigned to variant via date of symptom onset | | | | | | |
| *Alpha (N=546)* | Gamma | shape = 3.71 (0.31), rate = 0.50 (0.049) | 7.4 | 3.8 | 1.9 - 16.5 | 769 |
| | Weibull | scale = 8.45 (0.29), shape = 2.21 (0.12) | 7.5 | 3.6 | 1.5 - 15.4 | 763.1 |
| *Delta (N=286)* | Gamma | shape = 4.75 (0.64), rate = 0.76 (0.12) | 6.2 | 2.9 | 1.9 - 12.8 | 367.7 |
| | Weibull | scale = 7.04 (0.32), shape = 2.51 (0.19) | 6.3 | 2.7 | 1.6 - 12.0 | 373.3 |

## 2. Imputation of dates of infection

The task of reconstructing transmission chains must overcome the intrinsic limitation of the unobservability of transmission chains. We use available evidence to probabilistically impute plausible infection dates for all SARS-CoV-2 cases in our dataset. We combine observed dates of symptom onset, diagnosis, and negative test results with available knowledge on incubation periods and the probability of testing positive over time for infected individuals.

First, we impute the dates of infection for all symptomatic cases. Let $T_D$ be the date of diagnosis (when the individual tested positive), $T_{N,n}$ the date of the n-th negative test before diagnosis, and $T_S$ the date of symptom onset; we define the following probability $P_I$ of being infected on day $T_I$:

$$P_I(T_I) = f(T_D - T_I) \cdot \prod_n [1 - f(T_{N,n} - T_I)] \cdot P_S(T_s - T_I) \qquad \text{(Eq. 1)}$$

Where $f(t)$ is the probability of a SARS-CoV-2 case of testing positive after a time $t$ since infection and $P_S(t)$ is the probability density function of the incubation period. For f($t$), we use a previously estimated piecewise logistic function with one breakpoint [S1]. For $P_S(t)$ we use the average variant-specific estimate from contact tracing data in Reggio Emilia defined by the algorithm above as a baseline, and two previous alternative estimates on ancestral lineages [S2, S3] as sensitivity analyses (see Section S5-b and S5-c). For each symptomatic case, a time of infection is sampled from $P_I(t)$ and the date of infection $T_I$ is obtained by rounding to the closest integer; note that this sampling allows for possible false negative results in dates $T_{N,n}$. The sample is repeated K = 100 times.

For asymptomatic cases, we cannot use the information on the incubation period given that no date of symptom onset is defined. Therefore, we use the imputed dates of infection for symptomatic cases to define a distribution of diagnostic delays $P_D(x)$, defining the probability of being diagnosed after $x$ days from infection. An empirical approximation of $P_D(x)$ will be given, for any $x$, by the fraction of all instances across the K stochastic samples for which the diagnostic delay $T_R = T_D - T_I$ is equal to $x$. A gamma function is then fitted to the empirical distribution using a maximum likelihood approach to obtain $P_D(x)$. The infection date of asymptomatic cases can then be sampled from the following probability

$$P_I(T_I) = f(T_D - T_I) \cdot \prod_n [1 - f(T_{N,n} - T_I)] \cdot P_D(T_D - T_I) \qquad \text{(Eq. 2)}$$

assuming that the distribution of diagnostic delays for asymptomatic cases is the same as for symptomatic cases. Because this assumption cannot be tested, we use as a sensitivity analysis an alternative method where only the probabilities of negative and positive tests are used to define the $P_I$ (see Section S5-a). The sampling of infection times is repeated K times also for asymptomatic cases.

Assuming the imputation of incubation periods is correct, we obtain that 10.5% (95%CrI: 10.4-10.6%) of negative tests is a false negative result for the Alpha variant and 12.4% (95%CrI: 12.3-12.6%) for the Delta variant. This result was used to define the criterium of inclusion for households where undiagnosed cases have at least two negative tests, in order to reduce the fraction of undiagnosed positive cases to negligible levels (1.1% for Alpha and 1.5% for Delta) for the purpose of this analysis.

Figure S4 reports the estimated empirical and fitted distributions of diagnostic delays for variants Alpha and Delta.

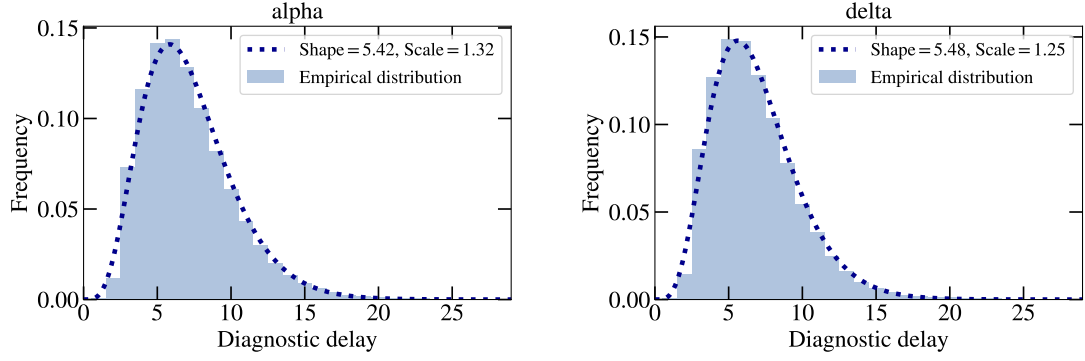

**Figure S4. Empirical and fitted distribution of the diagnostic delay P$_D$, estimated from symptomatic cases.** Left: Alpha variant; Right: Delta variant. The histograms represent the empirical distribution given the imputed infection times for symptomatic individuals. The dotted curves represent fitted gamma functions.

## 3. Inference of transmission links

The model adopted in this work extends the approach previously proposed in [S4]. We assumed that, at any time t, a susceptible individual j within a household is exposed to a force of infection composed of two components:

$$\lambda_j(t) = \lambda_j^o(t) + \lambda_j^h(t) \quad \text{(Eq. 3)}$$

Where $\lambda_j^o(t)$ represents the force of infection from the general community outside the household, and $\lambda_j^h(t)$ represents the one from infected members inside the household.
We define:

$$\lambda_j^o(t) = \sum_{z \in 0..t} \alpha I(z)\chi_j(z)\Gamma(t-z; a, b)q_j(t) \quad \text{(Eq. 4)}$$

Where:
- $\alpha$ is a free parameter scaling the transmissibility in the general community;
- $I(z)$ is proportional to the number of newly infected cases at time $z$ outside the household of $j$, obtained from epidemic curves by date of symptom onset for the province of Reggio Emilia in the Italian integrated surveillance system [S5, S6];
- $\chi_j(t)$ represents the relative susceptibility of individual $j$ and changes over time $t$ depending on the dates of vaccination of $j$;
- $\Gamma(t; a, b)$ represents the distribution of the intrinsic generation time at day t after infection, for which we assumed a discretized Gamma distribution with scale $a$ and shape $b$; in particular, given $g(t; a, b)$ the continuous Gamma probability distribution, $\Gamma(t; a, b) = \int_t^{t+1} g(\tau; a, b)d\tau$.
- q$_j$(t) is an on/off function that is 0 when the household of $j$ is in quarantine and 1 otherwise. For each household, a quarantine of 14 days is started after the first diagnosis and reinstated for a further 14 days every time there is a new diagnosis after the previous quarantine has ended.

In addition, we define $\lambda_j^h(t)$ as:

$$\lambda_j^h(t) = \sum_{i \in H_j} \lambda_{j,i}^h(t) = \sum_{i \in H_j} \beta \rho_i(t)\chi_j(t)\Gamma(t - T_{I,i}; a, b) \quad \text{(Eq. 5)}$$

where:
- $\rho_j(t)$ represents the relative transmissibility of individual $j$, and changes over time $t$ depending on the dates of vaccination of $j$;

- $i$ is an index running over the set $H_j$ of infected household members of individual $j$;
- $\beta$ is a free parameter scaling the transmissibility inside households;

For the relative susceptibility and we assumed that a vaccine dose starts to be protective 14 days after inoculation:

$$\chi_j(t) = \begin{cases} 1 & \text{if } t < t_{v,1} + 14 \\ 1 - \chi^{(1)} e^{-w(t-t_{v,1}-14)} & \text{if } t_{v,1} + 14 \leq t < t_{v,2} + 14 \\ 1 - \chi^{(2)} e^{-w(t-t_{v,2}-14)} & \text{if } t_{v,2} + 14 \leq t < t_{v,3} + 14 \\ 1 - \chi^{(3)} & \text{if } t \geq t_{v,3} + 14 \end{cases} \quad \text{(Eq. 6)}$$

Where $t_{v,d}$ is the date of vaccination dose $d$, $\chi^{(d)}$ are the initial effectiveness of dose $d$ (i.e., 14 days after vaccination) against the considered variant, and $w$ is the waning rate of vaccine protection. Estimates of vaccine effectiveness and waning rate were obtained from a large-scale retrospective cohort study on the Italian population [S7] and reported in Table S3.

**Table S3.** Parameters for vaccine effectiveness and waning.

| Parameter | Unit | Alpha | Delta |
| --- | --- | --- | --- |
| Initial effectiveness of dose 1 $\chi^{(1)}$ | % | 49.2 | 49.4 |
| Initial effectiveness of dose 2 $\chi^{(2)}$ | % | 81.9 | 80.2 |
| Waning rate $w$ | days$^{-1}$ | 0 | 1/227 |

Note that although there is high uncertainty on the effectiveness of the booster dose, only 3 individuals (all in the Delta period) had a booster dose within the study period.

For the relative transmissibility, we assumed a reduction by $\rho = 50\%$ after 14 days from the first dose [S8, S9]:

$$\rho_i(t) = \begin{cases} 1 & \text{if } t < t_{v,1} + 14 \\ \rho & \text{if } t \geq t_{v,1} + 14 \end{cases} \quad \text{(Eq. 7)}$$

The model assigns a source of infection $k_j$ for all cases by choosing from either a generic source outside the household or from an infectious household member in $H_j$, with probability proportional to the contribution of each source to the total force of infection $\lambda_j(T_{I,j})$ at the time $T_{I,j}$ at which j was infected. The overall likelihood of the observations given parameter set $\theta = (\alpha, \beta, a, b)$ and the assigned sources of infection $k_j$ is given by:

$$L(\theta, k_j) = \prod_j P_j Q_j \quad \text{(Eq. 8)}$$

where

$$P_j = \begin{cases} \lambda_j^o(T_{I,j}) & \text{if } k_j \text{ is outside the household} \\ \lambda_{j.i}^h(T_{I,j}) & \text{if } k_j \text{ is household member } i \\ 1 & \text{if } j \text{ is uninfected} \end{cases} \quad \text{(Eq. 9)}$$

For infected individuals, $Q_j$ is the probability that $j$ has not been infected until $t_{I,j}$, namely $Q_j = e^{-\int_0^{t_{I,j}} \lambda_j(t)dt}$. For uninfected individuals, it is the probability that $j$ has never been infected, $Q_j = e^{-\int_0^{\infty} \lambda_j(t)dt}$.

We estimated the unknown parameters $\theta$ and the source of infection $k_j$ for all cases using a Monte Carlo Markov Chain (MCMC) procedure. At each step, all parameters in $\theta$ are updated using reversible normal jumps. Z=500 samples from the posterior distributions obtained by the MCMC for each of the K=100 samples were pooled together to obtain the final parameter distribution and the distribution of the sources of infection for each case.

## 4. Additional results of the baseline model

*Statistics on reconstructed transmission links*

The average per-household number of infections contracted from the general community was 1.18 (95%CrI 1.16 – 1.22) during the Alpha period and 1.11 (95%CrI 1.08 – 1.14) during the Delta period. The average number of secondary infections generated by a positive case was 0.64 (0.63 – 0.65) during the Alpha period and 0.60 (0.59 – 0.61) during the Delta period. Table S4 shows how the model reconstructed transmission links within households with different numbers of cases.

**Table S4. Statistics for the model-based reconstruction of transmission links in households by number of SARS-CoV-2 cases.**

|  | Alpha | | Delta | |
|---|---|---|---|---|
|  | *Number* | *%* | *Number* | *%* |
| **Households with 2 SARS-CoV-2 cases** | **259** | **100** | **277** | **100** |
| - Both infected in the general community | 26 (18-36) | 10 (7-14) | 21 (12-29) | 7 (4-10) |
| - One infected the other | 232 (223-241) | 90 (86-94) | 256 (248-265) | 93 (90-96) |
| **Households with 3 SARS-Cov-2 cases** | **313** | **100** | **135** | **100** |
| - All infected in the general community | 2 (0-4) | 0.5 (0-1) | 0 (0-2) | 0 (0-1) |
| - One transmission, 2 infected in the general community | 43 (31-55) | 14 (10-18) | 15 (9-22) | 12 (7-17) |
| - Two transmissions, same infector (1 generation) | 114 (99-131) | 37 (32-42) | 54 (43-65) | 40 (32-48) |
| - Two transmissions, different infectors (2 generations) | 154 (136-172) | 49 (43-55) | 65 (54-77) | 48 (40-57) |
| **Households with 4 or more SARS-Cov-2 cases** | **334** | **100** | **112** | **100** |
| - All infected in the general community | 0 (0-1) | 0 (0-0) | 0 (0-0) | 0 (0-0) |
| - One transmission | 36 (25-47) | 11 (7-14) | 15 (8-21) | 13 (7-19) |
| - Two transmissions | 161 (145-177) | 48 (43-53) | 60 (50-69) | 53 (45-62) |
| - Three or more transmissions | 137 (108-172) | 41 (32-52) | 38 (24-54) | 33 (21-48) |

*Stability of the attributed source of infection*

For each case, we considered the distribution of the sources of infection attributed by the model and evaluated its stability. We categorized cases according to whether its source of infection was consistently (i.e., more than 75% of the times over the Z sampling of infector and K sampling of infectious dates) attributed to:
- the same household member;
- transmission within household but from different potential infectors;
- transmission in the general community.

The setting of transmission was uncertain (less than 75% consistency in attribution) in about 40% of cases (Figure S5) in both the Alpha and Delta periods. This generally happened when two or more cases in a household had close diagnosis dates, so that either could have been infected in the general community and then transmitted to the other, or both could have been infected in the general community, depending on the assigned dates of infection.

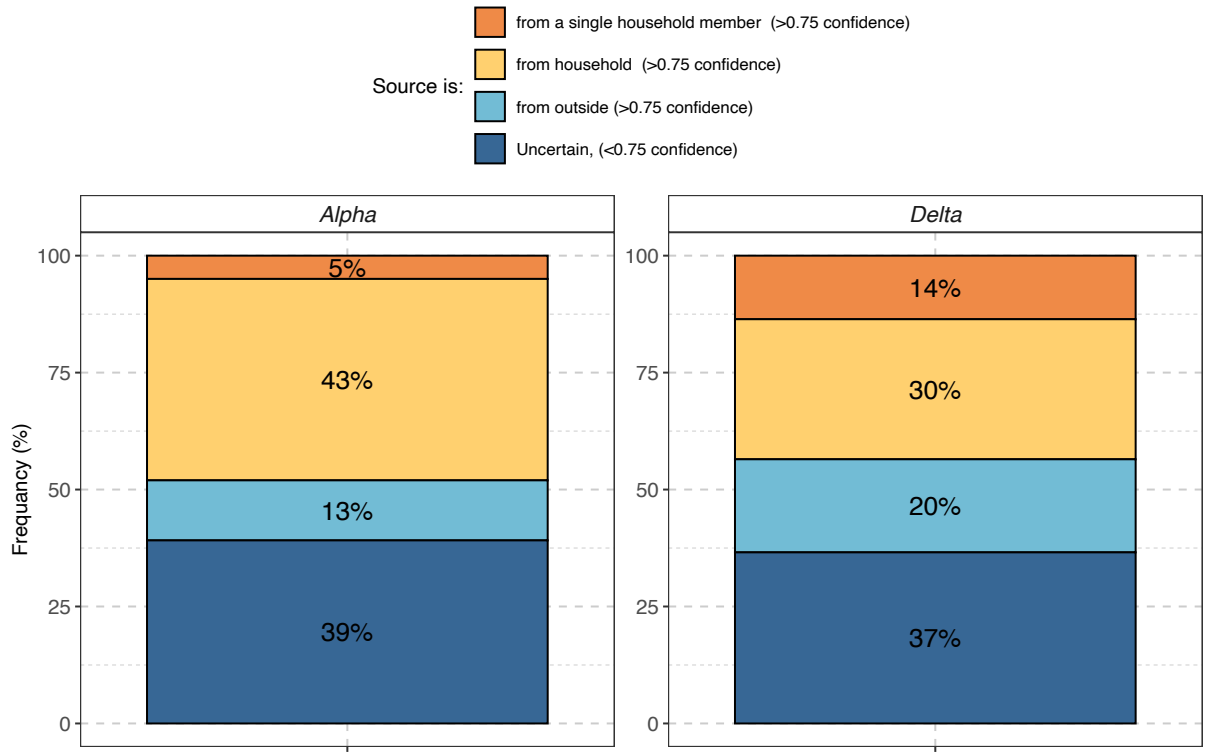

**Figure S5. Consistency in the attribution of the infector or the infector setting.** The stacked barchart represents the proportion of individuals that were consistently (more than 75% of the times across Z sampling of sources and K sampling of infectious dates) or inconsistently attributed to either category.

## 5. Sensitivity Analyses

We performed six sensitivity analyses (SA) to test the robustness of model results against different model assumptions. The first three SA (a-c) impact on the main unknown of the data, i.e. the imputed infectious periods of cases; the fourth (d) considers a reduced transmissibility for asymptomatic individuals; the fifth (e) evaluates the possibility that a fraction of undiagnosed individuals were fully protected from infection from previous natural immunity: the sixth (f) assumes that any effort to quarantine positive cases would not impact the force of infection from outside the household (*ie* $q(t) = 1$ for any value of t in equation 4)

## a) Imputation of dates of infection in asymptomatic cases

In the baseline method for the imputation of dates of infection, we implicitly assumed that symptomatic and asymptomatic cases have the same diagnostic delay distribution. We assess the impact of this assumption by considering an alternative method where the date of infection of asymptomatic individuals was assigned only on the basis of information on diagnostic date and negative test results. In this additional procedure the factor depending on $P_D$ is removed from Equation 2, resulting in:

$$P(j) = f(t_D - j) \cdot \prod_n [1 - f(n - j)] \qquad \text{(Eq. 10)}$$

Table S5 and Figure S6 show that results obtained in this sensitivity analysis are in line with the baseline.

**Table S5. Estimates for the intrinsic and realized generation time and serial intervals using an alternative method for the imputation of infection dates for asymptomatic individuals.**

|  |  | ALPHA | DELTA |
|---|---|---|---|
| **INTRINSIC GENERATION TIME** | mean (95%CrI) [days] | 6.71 (6.24-7.22) | 7.28 (6.41-8.28) |
|  | shape mean (95%CrI) | 2.46 (2.22-2.71) | 2.16 (1.87-2.47) |
|  | scale mean (95%CrI) | 2.74 (2.41-3.10) | 3.40 (2.74-4.32) |
| **REALIZED GENERATION TIME** | mean (95%CrI) [days] | 4.71 (4.70-4.75) | 4.31 (4.26-4.33) |
| **SERIAL INTERVAL** | mean (95%CrI) [days] | 2.55 (2.32-2.75) | 2.58 (2.40-2.75) |

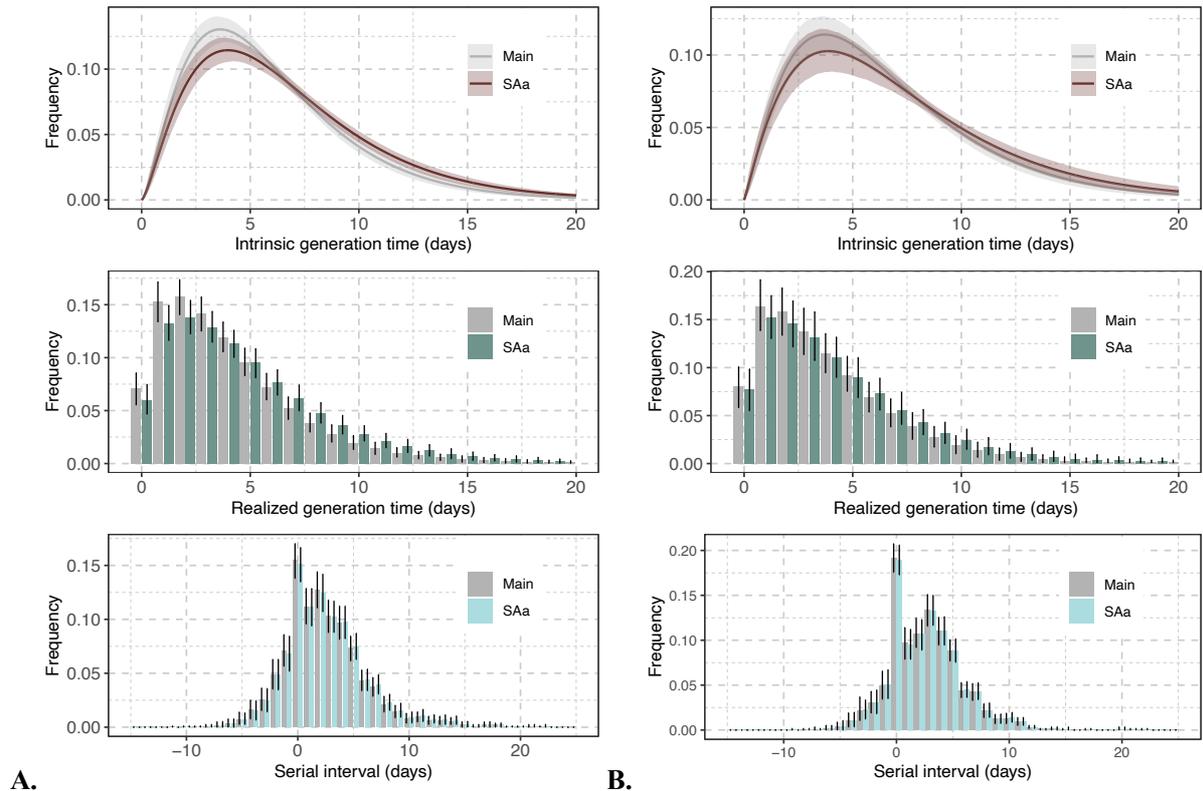

**Figure S6. Comparison between baseline analysis and results obtained with sensitivity analysis a), using an alternative method for the imputation of infection dates for asymptomatic individuals. A.** Alpha variant. **B.** Delta Variant.

## b) Distribution of the incubation period – I

In this sensitivity analysis, we reassigned infectious dates according to the baseline method, but using a different probability density function of the incubation period $P_S$ in Equation 1. We considered a gamma-distributed estimate for $P_S$ with shape 2.08 and scale 3.03 as derived for ancestral lineages in [S2]. Table S6 and Figure S7 show that results obtained in this sensitivity analysis are in line with the baseline.

**Table S6. Estimates for the intrinsic and realized generation time and serial intervals using an alternative distribution of incubation periods estimated for ancestral lineages in [S2].**

|  |  | **ALPHA** | **DELTA** |
|---|---|---|---|
| **INTRINSIC GENERATION TIME** | mean (95%CrI) [days] | 6.36 (5.98-6.76) | 7.76 (6.96-8.64) |
|  | shape mean (95%CrI) | 2.53 (2.29-2.82) | 2.16 (1.90-2.44) |
|  | scale mean (95%CrI) | 2.52 (2.19-2.85) | 3.62 (2.96-4.30) |
| **REALIZED GENERATION TIME** | mean (95%CrI) [days] | 4.41 (4.40-4.43) | 4.55 (4.51-4.56) |
| **SERIAL INTERVAL** | mean (95%CrI) [days] | 2.16 (1.92-2.37) | 2.13 (1.90-2.37) |

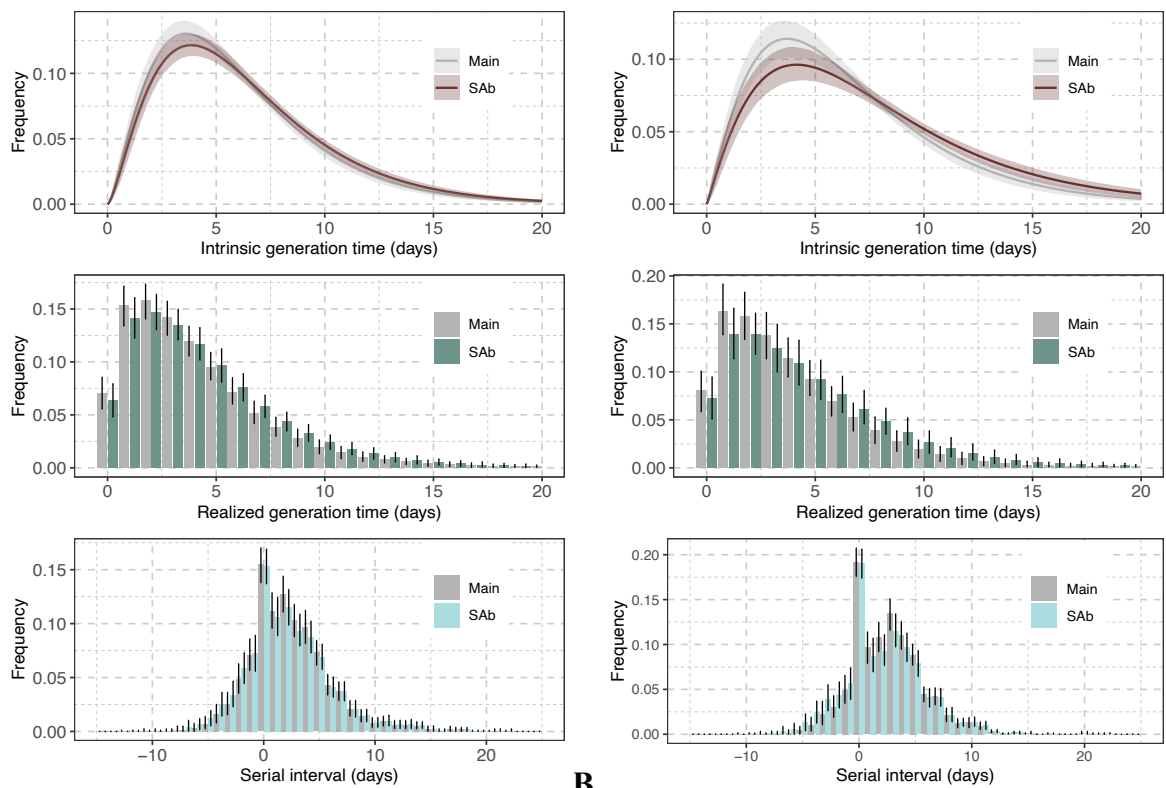

**Figure S7. Comparison between baseline analysis and results obtained with sensitivity analysis b), using an alternative distribution of incubation periods estimated for ancestral lineages in [S2]. A.** Alpha variant. **B.** Delta Variant.

## c) Distribution of the incubation period – II

Similarly to SA b), we considered a further alternative for the gamma-distributed estimate for Ps with shape 4.23 and scale 1.23, as derived for ancestral lineages in [S3]. Table S7 and Figure S8 show that results obtained in this sensitivity analysis are in line with the baseline.

**Table S7. Estimates for the intrinsic and realized generation time and serial intervals using an alternative distribution of incubation periods estimated for ancestral lineages in [S3].**

|  |  | ALPHA | DELTA |
|---|---|---|---|
| **INTRINSIC GENERATION TIME** | mean (95%CrI) [days] | 5.90 (5.58-6.24) | 6.93 (6.26-7.65) |
|  | shape mean (95%CrI) | 2.55 (2.33-2.82) | 2.23 (1.95-2.53) |
|  | scale mean (95%CrI) | 2.32 (2.03-2.60) | 3.13 (2.58-3.81) |
| **REALIZED GENERATION TIME** | mean (95%CrI) [days] | 4.05 (4.03-4.06) | 4.09 (4.07-4.11) |
| **SERIAL INTERVAL** | mean (95%CrI) [days] | 2.49 (2.31-2.66) | 2.45 (2.25-2.64) |

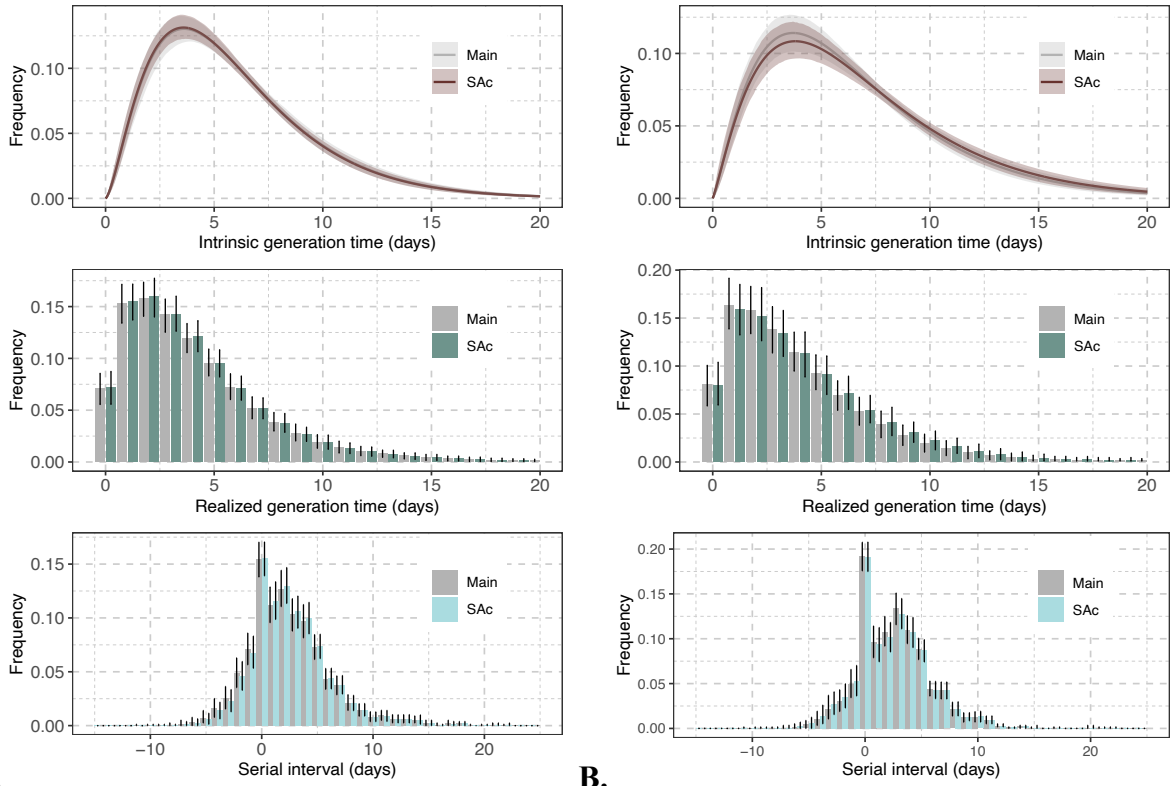

**Figure S8. Comparison between baseline analysis and results obtained with sensitivity analysis c), using an alternative distribution of incubation periods estimated for ancestral lineages in [S3]. A.** Alpha variant. **B.** Delta Variant.

*d) Reduced transmissibility for asymptomatic individuals*

In this sensitivity analysis, we consider a halved transmissibility for asymptomatic individuals [S10] by modifying Equation 7 as follows:

$$\rho_i(t) = \begin{cases} \varphi_i & if\ t < t_{v,1} + 14 \\ \rho\varphi_i & if\ t \geq t_{v,1} + 14 \end{cases} \quad \text{(Eq. 11)}$$

Where $\varphi_i$ is 1 if $i$ is symptomatic and 0.5 if asymptomatic. Table S8 and Figure S9 show that results obtained in this sensitivity analysis are in line with the baseline.

**Table S8. Estimates for the intrinsic and realized generation time and serial intervals using a halved transmissibility for asymptomatic individuals.**

|  |  | ALPHA | DELTA |
|---|---|---|---|
| **INTRINSIC GENERATION TIME** | mean (95%CrI) [days] | 6.51 (6.07-7.03) | 6.96 (6.30-7.84) |
|  | shape mean (95%CrI) | 2.56 (2.23-2.91) | 2.28 (1.97-2.58) |
|  | scale mean (95%CrI) | 2.56 (2.17-3.05) | 3.08 (2.53-3.83) |
| **REALIZED GENERATION TIME** | mean (95%CrI) [days] | 4.26 (4.22-4.28) | 4.02 (4.00-4.05) |
| **SERIAL INTERVAL** | mean (95%CrI) [days] | 2.54 (2.35-2.72) | 2.59 (2.41-2.76) |

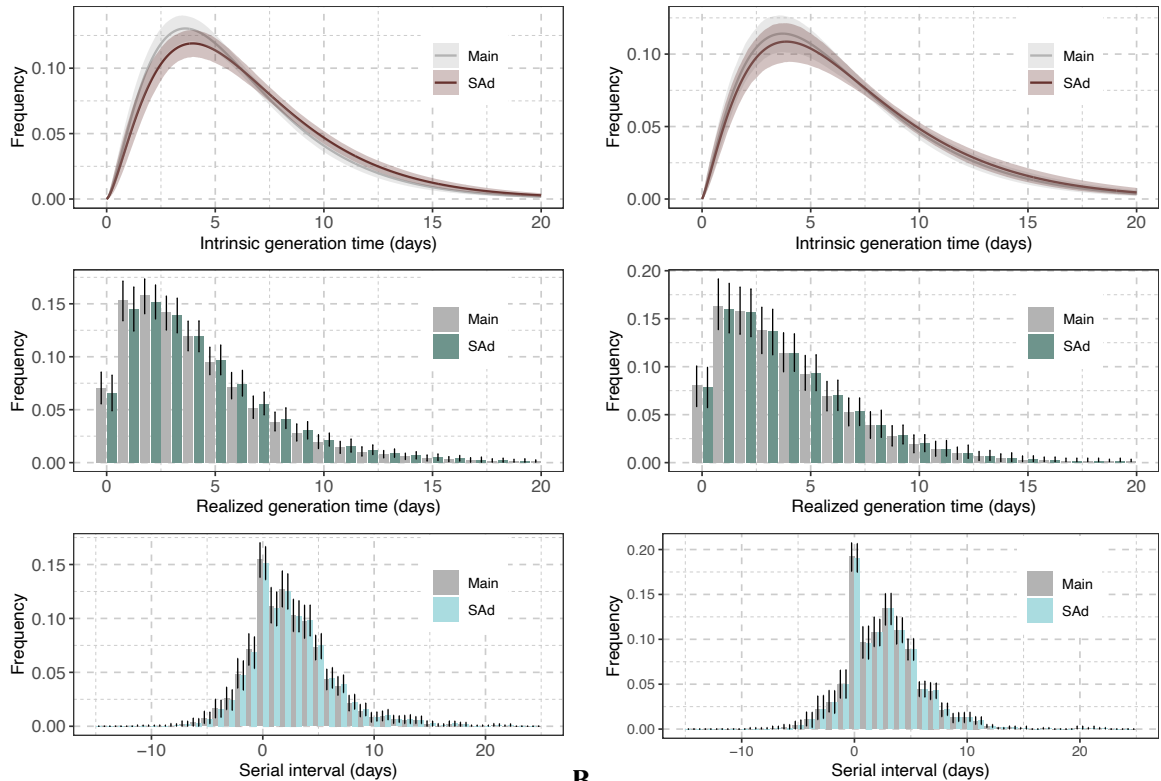

**A. B.**
**Figure S9. Comparison between baseline analysis and results obtained with sensitivity analysis d), using a halved transmissibility for asymptomatic individuals. A.** Alpha variant. **B.** Delta Variant.

*e) Protection from previous infection in a fraction of undiagnosed household members*

In this sensitivity analysis, we assume that a fraction of individuals who were undiagnosed were not susceptible to infection due to immunity conferred by previous SARS-CoV-2 infection. Using previous estimates of the cumulative SARS-CoV-2 attack rate in Italy before the Alpha and the Delta waves [S11], we assume that 15% of undiagnosed household cases during the Alpha period and 20% of undiagnosed household cases during the Delta period were immune. These cases were randomly sampled and removed from set of *j* for each of the Z repetitions of the MCMC procedure. The absence of these cases impacts on the component of $Q_j$ of the likelihood in Equation 8. Table S9 and Figure S10 show that results obtained in this sensitivity analysis are in line with the baseline.

**Table S9. Estimates for the intrinsic and realized generation time and serial intervals when assuming that 15% of undiagnosed cases in the Alpha period and 20% of undiagnosed cases in the Delta period were protected from infection via natural immunity from previous infection.**

|  |  | ALPHA | DELTA |
|---|---|---|---|
| **INTRINSIC GENERATION TIME** | mean (95%CrI) [days] | 6.12 (5.74-6.57) | 7.00 (6.35-7.78) |
|  | shape mean (95%CrI) | 2.56 (2.32-2.79) | 2.25 (2-2.54) |
|  | scale mean (95%CrI) | 2.40 (2.12-2.69) | 3.13 (2.56-3.79) |
| **REALIZED GENERATION TIME** | mean (95%CrI) [days] | 4.12 (4.1-4.14) | 4.00 (3.98-4.04) |
| **SERIAL INTERVAL** | mean (95%CrI) [days] | 2.43 (2.23-2.62) | 2.58 (2.39-2.76) |

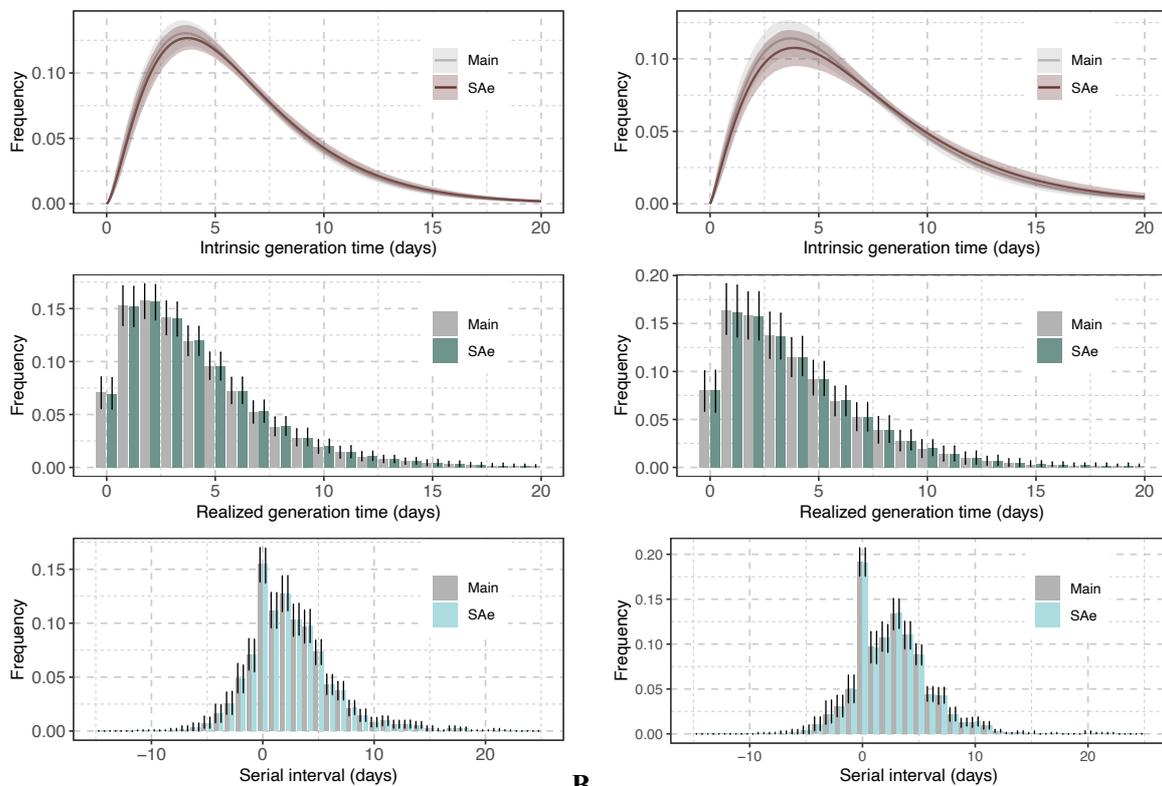

**Figure S10. Comparison between baseline analysis and results obtained with sensitivity analysis e), assuming that 15% of undiagnosed cases in the Alpha period and 20% of undiagnosed cases in the Delta period were protected from infection via natural immunity from previous infection. A.** Alpha variant. **B.** Delta Variant.

*f) No protection from infection outside the household during quarantine*

In this sensitivity analysis, we assume that the imposed quarantine period after the first positive diagnosis would not impact the force of infection from outside the household (i.e., $q(t) = 1$ for any value of t in Equation 4. Table S10 and Figure S11 show that results obtained in this sensitivity analysis are in line with the baseline.

**Table S10. Estimates for the intrinsic and realized generation time and serial intervals when assuming no protection from outside infection during the quarantine period.**

|  |  | ALPHA | DELTA |
|---|---|---|---|
| **INTRINSIC GENERATION TIME** | mean (95%CrI) [days] | 5.42 (4.96-5.83) | 6.4 (5.74-7.13) |
|  | shape mean (95%CrI) | 2.78 (2.53-3.06) | 2.35 (2.07-2.71) |
|  | scale mean (95%CrI) | 1.96 (1.68-2.24) | 2.74 (2.19-3.36) |
| **REALIZED GENERATION TIME** | mean (95%CrI) [days] | 3.84 (3.81-3.89) | 3.89 (3.87-3.96) |
| **SERIAL INTERVAL** | mean (95%CrI) [days] | 2.24 (2.04-2.45) | 2.52 (2.33-2.7) |

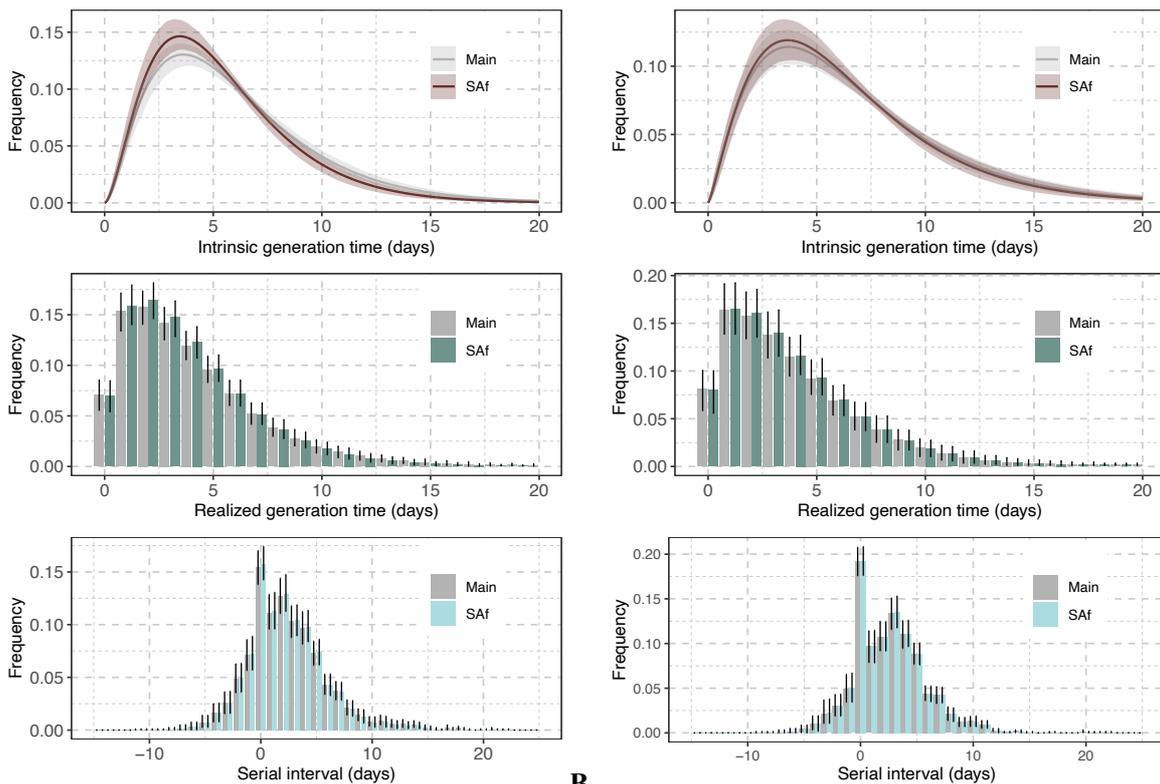

A.    B.

**Figure S11. Comparison between baseline analysis and results obtained with sensitivity analysis f), assuming no protection from outside infection during the quarantine period. A.** Alpha variant. **B.** Delta Variant.